\documentclass[twocolumn,showpacs,preprintnumbers,amsmath,amssymb,pra]{revtex4}

\usepackage{color}

\usepackage{graphicx}
\usepackage{dcolumn}
\usepackage{bm}
\usepackage{epstopdf}
\usepackage{bbold}
\usepackage{amsthm}
\newtheorem{theorem}{Theorem}

\newcommand{\iden}{\mathbb{1}}

\begin{document}

\preprint{}

\title{Channels that do not generate coherence}

\author{Xueyuan Hu}
\email{xyhu@sdu.edu.cn}
\affiliation{School of Information Science and Engineering, and Shandong Provincial Key Laboratory of Laser Technology and Application, Shandong University, Jinan 250100, China}

\date{\today}

\begin{abstract}
We define the coherence non-generating channel as the completely positive trace-preserving map which does not generate quantum coherence from an incoherent state. The incoherent operations are the strict subset of the non-coherence-generating channels. Although the relative entropy of coherence is monotonically decreasing under the non-coherence-generating channels, we prove that the coherence of formation may increase under such channels. Interestingly, by building a mathematical relation between the coherence of formation and the entanglement of formation, we show that the coherence of formation of a single-qubit state is never increased by a non-coherence-generating channel. This leads to the superadditivity property for the coherence increasing power of quantum channels, namely, while two channels can not increase coherence individually, they may increase the quantum coherence of a composed system. Further, we derive the general form of the rank-2 coherence non-generating qubit channels. Our results contribute to the resource theory of quantum coherence.
\end{abstract}

\maketitle

\section{Introduction}
Although quantum resources, such as entanglement \cite{RevModPhys.81.865}, quantum correlations \cite{RevModPhys.84.1655}, quantum steering \cite{PhysRevX.5.041008}, etc., are of fundamental importance for quantum information processing, it is only recently that the general framework of quantum resource theory has been built \cite{Horo_Re_Th,PhysRevLett.115.070503}. Underlying a quantum resource theory, there are three basic ingredients: the free operations (which can be implemented at no cost during a quantum information task), the free states (which can be prepared using the free operations), and the resource states (which are prepared before and can be used as a resource in the task). Apparently, one can never prepare a resource state from a free state using the free operations. However, one can use the free operations to prepare several copies of a resource state from more copies of another state with less resource, or from less copies of another state with more resource. An important result has been proved\cite{PhysRevLett.115.070503}, namely, that if the free operations are the maximal set of operations which do not generate resource from a free state, the state transformation is reversible. Here the reversibility means that, if $n$ copies of state $\rho$ can be transformed to $m$ copies of state $\sigma$ using the free operations, one can also retrieve $n$ copies of $\rho$ from $m$ copies of $\sigma$, in the limit of $\max\{n,m\}\rightarrow\infty$.

The superpositions of quantum states, or the quantum coherence \cite{PhysRevLett.113.140401,arXiv:1506.07975}, serve as a ``resource'' in quantum information tasks such as quantum algorithms \cite{INSPEC:4865020} and quantum key distribution \cite{PhysRevLett.68.3121}. Further, the quantum coherence has been related to other well-studied quantum resources such as entanglement \cite{PhysRevLett.115.020403}, quantum correlations \cite{arXiv:1507.08171, arXiv1506.01773, arXiv:1508.01978, arXiv:1510.06179}, and randomness \cite{PhysRevA.92.022124}. In the resource theory of quantum coherence, the free states are the incoherent states, the density matrices of which are diagonal on the reference basis. The set of incoherent states is denoted as $\mathcal I$. The free operations are the incoherent operations (ICs) $\Lambda^{\mathrm{IC}}$, for which there exists a Kraus decomposition $\Lambda^{\mathrm{IC}}(\cdot)=\sum_nK_n(\cdot)K_n^\dagger$ such that $\rho_n\equiv\frac{K_n\rho K_n^\dagger}{\mathrm{tr}(K_n\rho K_n^\dagger)}$ is incoherent for any incoherent state $\rho$. The set of incoherent operations is labeled as $\mathcal{IC}$. Any incoherent operation can be implemented in the following way: one applies a unitary operator $U$ to the particle and an ancilla $A$ such that $U\rho\otimes|0\rangle_A\langle0|U^\dagger=\sum_nK_n\rho K_n^\dagger\otimes|n\rangle\langle n|\in\mathcal I,\ \forall\ \rho\in\mathcal I$, and then throws the ancilla away. During the process, no coherence is produced in the composed system. However, for a channel $\Lambda^C\notin\mathcal{IC}$, coherence must be generated in the composed system. In this sense, we say that the channels not in $\mathcal{IC}$ can not be implemented incoherently. The resource theory of coherence has been proved irreversible for general states, so the incoherent operations are a strict subsect of non-coherence-generating channels (NCs). It is then of interest to study the formation and properties of non-coherence-generating channels, especially those not belong to the set of incoherent operations.

In this paper, we consider the whole set of quantum channels which never generate coherence from any incoherent state. Like ICs, the non-coherence-generating channels never increase the coherence of entropy $C_r$ of any input state. 
Despite the monotonicity of the coherence of formation $C_f$ under ICs, we prove that $C_f$ is not monotonically decreasing under NCs. Interestingly, while $C_f$ of a single qubit is proved never increased by NCs, we present an example where the coherence of the two-qubit state is increased by a local NC. Further, this superadditivity property of the coherence increasing power is proved for any two channels. Besides, we derive the general form of rank-2 qubit NCs.

\section{Quantum coherence measures}
In order to quantify the quantum coherence, we employ the relative entropy of coherence $C_r$ and the coherence of formation $C_f$, which are defined as \cite{arXiv:1506.07975}
\begin{eqnarray}
\label{eq:cr}C_r(\rho)&:=&\min_{\sigma\in\mathcal I} S(\rho||\sigma)=S\bm(\Delta(\rho)\bm)-S(\rho),\\
\label{eq:cf}C_f(\rho)&:=&\min_{\{p_i,|\psi_i\rangle\}}\sum_ip_iS\bm(\Delta(\psi_i)\bm).
\end{eqnarray}
Here, $S(\rho)=-\mathrm{tr}(\rho\log_2\rho)$ is the von Neumann entropy, $S(\rho||\sigma)=\mathrm{tr}[\rho(\log_2\rho-\log_2\sigma)]$ is the relative entropy, $\{p_i,|\psi_i\rangle\}$ is a pure state decomposition of state $\rho$, $\psi_i\equiv|\psi_i\rangle\langle\psi_i|$ is called a pure state component of $\rho$, and the decohering operation $\Delta$ erases all of the off-diagonal elements of the density matrix $\Delta(\rho)=\sum_i\langle i|\rho|i\rangle|i\rangle\langle i|$.

From Winter and Yang \cite{arXiv:1506.07975}, the measures $C_r$ and $C_f$ have certain properties, such as the following.\\
(C1) Both $C_r(\rho)$ and $C_f(\rho)$ vanish iff $\rho\in\mathcal I$, and reach unity when $\rho=\Phi_2\equiv|\Phi_2\rangle\langle\Phi_2|=\frac12\sum_{i,j=0}^{1}|i\rangle\langle j|$.\\
(C2) Both $C_r$ and $C_f$ are monotonically decreasing under incoherent operations.\\
(C3) Both $C_r$ and $C_f$ are additive, $C_r(\rho\otimes\sigma)=C_r(\rho)+C_r(\sigma)$ and $C_f(\rho\otimes\sigma)=C_f(\rho)+C_f(\sigma)$.\\
(C4) Operational interpretations. $C_r(\rho)=C_{dis}^{\mathrm{IC}}(\rho)$ and $C_f(\rho)=C_{cos}^{\mathrm{IC}}(\rho)$. Here the distillable coherence $C_{dis}^{\mathrm{IC}}(\rho)$ is the maximal asymptotic rate at which the unit coherence state $\Phi_2$ can be distilled from $\rho$ by IC, and the coherence cost $C_{cos}^{\mathrm{IC}}(\rho)$ is the minimal asymptotic rate of consuming $\Phi_2$ for preparing $\rho$ by IC.\\
(C5) $C_f(\rho)\geq C_r(\rho)$. The equality holds iff $\rho$ is in the form
\begin{equation}
\rho=\oplus_jp_j|\phi_j\rangle\langle\phi_j|,\label{re_state}
\end{equation}
where $|\phi_j\rangle$ are all supported on the orthogonal subspaces spanned by a partition of the incoherent basis.

For later convenience, we also mention a couple of coherence monotones defined on distances. The $l_1$-norm of coherence \cite{PhysRevLett.113.140401} is defined as the minimum $l_1$-norm distance from $\rho$ to the set of incoherent states, and happens to have the simple formula $C_{l_1}(\rho)=\sum_{i\neq j}|\langle i|\rho|j\rangle|$. The trace distance of coherence is defined as $C_{tr}(\rho):=\min_{\sigma\in\mathcal I}\|\rho-\sigma\|_{tr}$, where $\|\rho-\sigma\|_{tr}=\mathrm{tr}\sqrt{(\rho-\sigma)^\dagger(\rho-\sigma)}$ is the trace distance between $\rho$ and $\sigma$. As proved in \cite{PhysRevA.93.012110}, $C_{l_1}(\rho)=C_{tr}(\rho)$ if the density matrix $\rho$ is in a block diagonal form on the incoherent basis, where the dimension of each block is at most 2.

\section{Non-coherence-generating channels}
\textbf{Definition 1.} \emph{A non-coherence-generating channel $\Lambda^{\mathrm{NC}}$ is a completely positive trace-preserving map from an incoherent state to an incoherent state
\begin{equation}
\Lambda^{\mathrm{NC}}(\mathcal I)\subset\mathcal I.
\end{equation}}
The set of non-coherence-generating channels is denoted as $\mathcal{NC}$. Some direct properties of the non-coherence-generating channels are observed. \\
(P1) Because the resource theory of coherence is irreversible for general states, the incoherent operations are a strict subset of the non-coherence-generating channels: $\mathcal{IC}\subset\mathcal{NC}$. \\
(P2) The cohering power of a channel vanishes iff the channel is a non-coherence-generating channel. \\
(P3) The tensor produce of two non-coherence-generating channel is a non-coherence-generating channel: $\Lambda\otimes\mathcal E\in\mathcal{NC},\ \forall\ \Lambda,\mathcal E\in\mathcal{NC}$. \\
(P4) Two non-coherence-generating channels are composed to a non-coherence-generating channel: $\Lambda\circ\mathcal E\in\mathcal{NC},\ \forall\ \Lambda,\mathcal E\in\mathcal{NC}$.

\section{The (non)monotonicity of coherence measures under NC}
Since $C_r$ and $C_f$ has the operational interpretation of the distillable coherence and coherence cost, respectively, the monotonicities of $C_r$ and $C_f$ under certain set of quantum channels are directly related to the efficiency of coherence distillation and formation. Despite the monotonicity of both $C_r$ and $C_f$ under the incoherent operations, it is not \textit{a priori} clear whether they are still monotonic under the non-coherence-generating channels. Here we prove that $C_r$ is monotonically decreasing under NC, but the (non)monotonicity of $C_f$ is more complicated. Although for a single-qubit state, $C_f$ is proved never increased by any NC, $C_f$ of a higher-dimension state is generally \textit{not} monotonic under NC.

\begin{theorem}\label{mono_r}
The relative entropy of coherence $C_r$ never increases under any non-coherence-generating channel,
\begin{equation}
C_r(\Lambda^{\mathrm{NC}}(\rho))\leq C_r(\rho),\ \forall\ \Lambda^{\mathrm{NC}}\in\mathcal{NC},\rho.
\end{equation}
\end{theorem}
The proof is in Appendix A. A consequence of this theorem is that, the optimal asymptotic rate of distilling $\Phi_2$ from $\rho$ using the whole set of non-coherence-generating channels is the relative entropy of coherence,
\begin{equation}
C_{dis}^{\mathrm{NC}}(\rho)=C_r(\rho).\label{eq:dis}
\end{equation}
(See Appendix B for details.) This is consistent with the general resource theory. Actually, as we have considered the whole set of NCs, the transformations between coherent states become reversible; meanwhile, for any reversible quantum resource theory, the unique asymptotic rate is proved to be the regularized relative entropy of a resource \cite{PhysRevLett.89.240403}. Recalling $C_{dis}^{\mathrm{IC}}(\rho)=C_r(\rho)$, Eq. (\ref{eq:dis}) means that generalizing the allowed operations from $\mathcal{IC}$ to $\mathcal{NC}$ does not make the coherence distillation any more efficient.

The behavior of $C_f$ under NC is more complicated than $C_r$. Before dealing with the (non)monotonicity of $C_f$, we first prove two lemmas, which mathematically relate $C_f$ to the entanglement of formation $E_f$. For a bipartite state $\rho^{AB}$, the entanglement of formation is defined as
\begin{equation}
\label{eq:ef}E_f(\rho^{AB}):=\min_{\{p_k,|\Psi_k\rangle\}}\sum_kp_kS(\mathrm{tr}_A(\Psi_k)),
\end{equation}
where ${\{p_k,|\Psi_k\rangle\}}$ is a pure state decomposition of $\rho^{AB}$ and $\Psi_k\equiv|\Psi_k\rangle\langle\Psi_k|$. We observe the similarity of Eqs. (\ref{eq:cf}) and (\ref{eq:ef}), and prove Lemma 1 (see Appendix C for details).

\textbf{Lemma 1.} \emph{For any $d$-dimension state $\rho_d=\sum_{i,j=0}^{d-1}\rho_{ij}|i\rangle\langle j|$, there is a maximally correlated state $\rho_{d\times d}=\sum_{i,j=0}^{d-1}\rho_{ij}|ii\rangle\langle jj|$, whose entanglement of formation equals the coherence of formation of $\rho_d$ on the reference basis $\{|i\rangle\}$,
\begin{equation}
C_f(\rho_d)=E_f(\rho_{d\times d}).\label{coh_ent}
\end{equation}}

According to \cite{PhysRevLett.78.5022}, the entanglement of formation $E_f(\rho_{d\times d})$ for $d=2$ is $E_f(\rho_{2\times2})=h(\frac{1+\sqrt{1-\mathrm{Con}^2(\rho_{2\times2})}}{2})$, where the concurrence is calculated as $\mathrm{Con}(\rho_{2\times2})=2|\rho_{01}|=C_{l_1}(\rho_2)$. Hence we arrive at the following lemma.

\textbf{Lemma 2.} \emph{For a qubit state $\rho_2$, we have
\begin{equation}
C_f(\rho_2)=h(\frac{1+\sqrt{1-C^2_{l_1}(\rho_2)}}{2}),\label{cf2}
\end{equation}
where $h(x):=-x\log_2x-(1-x)\log_2(1-x),x\in[0,1]$.}

From Lemma 2, we observe that for a single-qubit state $\rho_2$, the coherence of formation $C_f(\rho_2)$ is monotonically increasing with $C_{l_1}(\rho_2)$. Further, it can be proved that $C_{l_1}(\rho_2)$ can never be increased by any coherence non-generating \emph{qubit} channel. Hence we arrive at Theorem \ref{mono_f2} (see Appendix D for details).

\begin{theorem}\label{mono_f2}
The coherence of formation for a single-qubit state $\rho_2$ can not be increased by non-coherence-generating channels,
\begin{equation}
C_f(\Lambda^{\mathrm{NC}}_2(\rho_2))\leq C_f(\rho_2),\ \forall\ \Lambda^{\mathrm{NC}}_2\in\mathcal{NC}_2,\rho_2\in\mathcal D(\mathcal H_2).\label{mono_f2_e}
\end{equation}
Here $\mathcal D(\mathcal H_2)$ denotes the set of density operator acting on the two-dimensional Hilbert space $\mathcal H_2$, and $\mathcal{NC}_2$ is a set of single-qubit NCs $\Lambda^{\mathrm{NC}}_2:\mathcal D(\mathcal H_2)\rightarrow\mathcal D(\mathcal H_2)$.
\end{theorem}

Although $C_f$ of a single-qubit state does not increase under the NC channels, the irreversibility of the coherence resource theory require that $C_f$ must not be always monotonically decreasing. This leads to the following theorem (see Appendix E for detailed proof).

\begin{theorem}\label{mono_f}
The quantum coherence of formation $C_f$ can be increased by some non-coherence-generating channels
\begin{equation}
C_f(\Lambda^{\mathrm{NC}}(\rho))> C_f(\rho),\ \exists\ \Lambda^{\mathrm{NC}}\in\mathcal{NC},\rho.
\end{equation}
\end{theorem}

Theorems \ref{mono_f2} and \ref{mono_f} do not conflict with each other. Although none of the qubit NC ever increases $C_f$ of a single-qubit state, it is still possible for some qubit NCs to increase $C_f$ when applied to each qubit of some multi-qubit states, i. e., $\exists\ \Lambda^i_2\in\mathcal{NC}_2,\rho\in\mathcal D(\mathcal H_2^{\otimes n})$ such that
\begin{equation}
C_f\left(\bigotimes_{i=1}^n\Lambda^i_2(\rho)\right)>C_f(\rho).\label{sup_ad}
\end{equation}
We name this property the superadditivity of the coherence increasing power, and study it explicitly in the next section.

\section{Coherence increasing power and its superadditivity}
From the above section, the quantum channels which never generate coherence from an incoherent state may still have power to increase $C_f$. Based on this observation, we define the coherence increasing power of qubit channels, which is completely different from the cohering power as defined in \cite{PhysRevA.92.032331}.

\textbf{Definition 2.} \emph{Let $\Lambda_d:\mathcal D(\mathcal H_d)\rightarrow\mathcal D(\mathcal H_d)$ be a quantum operation. The coherence increasing power of $\Lambda_d$ is defined as
\begin{equation}
P_{C}(\Lambda_{d})=\sup_{\rho\in\mathcal D(\mathcal H_d)}C(\Lambda_{d}(\rho))-C(\rho).
\end{equation}
Here the coherence measure $C$ can be chosen as $C_r$ or $C_f$.}

The non-negative function $P_C(\Lambda_d)$ vanishes for incoherent operations. $P_{C_r}$ also vanishes for all of the non-coherence-generating channels which are not incoherent, while $P_{C_f}$ can be positive for such channels.

Here we present an example where $P_{C_f}(\Lambda^1_2)=P_{C_f}(\Lambda^2_2)=0$ but $P_{C_f}(\Lambda^1_2\otimes\Lambda^2_2)>0$. Let $\Lambda^1_2=\iden_2$ and $\Lambda^2_2(\cdot)=E_1(\cdot)E_1^\dagger+E_2(\cdot)E_2^\dagger$ with
\begin{equation}
E_1=\frac{1}{2}\left(\begin{array}{cc}
1 & 0\\
-1 & \sqrt2
\end{array}\right),
E_2=\frac{1}{2}\left(\begin{array}{cc}
1 & \sqrt2\\
1 & 0
\end{array}\right).
\end{equation}
It can be checked that both of the qubit channels $\iden_2$ and $\Lambda^2_2$ are non-coherence-generating channels. From Theorem \ref{mono_f2}, $P_{C_f}(\iden_2)=P_{C_f}(\Lambda^2_2)=0$.

In order to show $P_{C_f}(\iden_2\otimes\Lambda^2_2)>0$, we only need to find a two-qubit state $\rho$ which satisfies Eq. (\ref{sup_ad}) with $n=2$. Here we choose $\rho=\Phi^+\equiv|\Phi^+\rangle\langle\Phi^+|$ with $|\Phi^+\rangle=\frac{1}{\sqrt2}(|00\rangle+|11\rangle)$. The output state is then a rank-2 state $\rho_{out}=\iden_2\otimes\Lambda^2_2(\Phi^+)=\frac12(|v_1\rangle\langle v_1|+|v_2\rangle\langle v_2|)$ with $|v_1\rangle=\frac{1}{\sqrt2}(\sin\frac{\pi}{8}|00\rangle+\cos\frac{\pi}{8}|01\rangle+\cos\frac{\pi}{8}|10\rangle-\sin\frac{\pi}{8}|11\rangle)$ and $|v_2\rangle=\frac{1}{\sqrt2}(\cos\frac{\pi}{8}|00\rangle-\sin\frac{\pi}{8}|01\rangle+\sin\frac{\pi}{8}|10\rangle+\cos\frac{\pi}{8}|11\rangle)$. The input state is a pure state, and simple calculation leads to $C_f(\Phi^+)=1$. The calculation of $C_f(\rho_{out})$ is complicated, but we prove (in Appendix F) that $C_f(\rho_{out})$ is strictly larger than 1. Therefore, $C_f(\iden_2\otimes\Lambda^2_2(\Phi^+))-C_f(\Phi^+)>0$, and hence $P_{C_f}(\iden_2\otimes\Lambda^2_2)>0$.

So far, we have presented an example where the coherence increasing powers of two qubit channels are superadditive. Next, we prove a general theorem of the superadditivity of the coherence increasing power, from the additivity of $C_r$ and $C_f$ under tensor products (see Appendix G for details).

\begin{theorem}
For any finite dimensional quantum channels $\Lambda_{d_1}$ and $\mathcal E_{d_2}$,
\begin{equation}
P_{C}(\Lambda_{d_1}\otimes\mathcal E_{d_2})\geq P_{C}(\Lambda_{d_1})+P_{C}(\mathcal E_{d_2}).
\end{equation}
\end{theorem}

The reason for the superadditivity of coherence increasing power is that, in the composed Hilbert space $\mathcal H_{d_1}\otimes \mathcal H_{d_2}$, the coherence of state does not only exhibit as local coherence but also the correlation between the two particles. Further, instead of the maximally correlated states, the maximally coherent states in $\mathcal H_{d_1}\otimes \mathcal H_{d_2}$ are the tensor product states of maximally coherent states in $\mathcal H_{d_1}$ and $\mathcal H_{d_2}$. This provides the opportunity to turn the quantum correlation into the local coherence, and meanwhile increase the quantum coherence of the total state. In other words, $\mathcal H_{d_1}\otimes \mathcal H_{d_2}$ provides a larger playground to exhibit the cohering property of $\Lambda_{d_1}$ and $\mathcal E_{d_2}$.

The superadditivity of $P_{C_f}$ provides a criterion to check whether a conherence non-generating qubit channel is an incoherent channel.

\section{The coherence non-generating qubit channels}
In the Bloch presentation, the action of a qubit channel $\Lambda$ on $\rho$ is equivalent to a matrix $[\lambda_{ij}]_{i,j=0}^3$ acting on the four-dimensional vector $\boldsymbol R=(1,r_1,r_2,r_3)^\mathrm{T}$. Here, $\lambda_{00}=1$ and $\lambda_{01}=\lambda_{02}=\lambda_{03}=0$ are satisfied to make sure that the channel is trace preserving.

A qubit state is incoherent if and only if it lies in the $z$ direction. Thus, a qubit NC acts as $\Lambda^{\mathrm{NC}}_{2}:(0,0,r_3)^\mathrm{T}\mapsto(0,0,r'_3)^\mathrm{T}$. The corresponding matrix elements for $\Lambda^{\mathrm{NC}}_{2}$ satisfy $\lambda_{10}=\lambda_{20}=\lambda_{13}=\lambda_{23}=0$. Based on this consideration, we derive a class of the coherence non-generating qubit channels. A rank-2 qubit channel is a NC if and only if it has the Kraus decomposition either as $\Lambda^1(\cdot)=E_1^1(\cdot)E_1^{1\dagger}+E_2^1(\cdot)E_2^{1\dagger}$ with
\begin{eqnarray}
 E^1_1&=&\left(\begin{array}{cc}
e^{i\eta}\cos\theta\cos\phi & 0\\
-\sin\theta\sin\phi & e^{i\xi}\cos\phi
\end{array}\right),\nonumber\\
E_2^1&=&\left(\begin{array}{cc}
\sin\theta\cos\phi & e^{i\xi}\sin\phi\\
e^{-i\eta}\cos\theta\sin\phi & 0
\end{array}\right).
\end{eqnarray}
or as $\Lambda^2(\cdot)=E_1^2(\cdot)E_1^{2\dagger}+E_2^2(\cdot)E_2^{2\dagger}$ with
\begin{eqnarray}
E_1^2=\left(\begin{array}{cc}
\cos\theta & 0\\
0 & e^{i\xi}\cos\phi
\end{array}\right),
E_2^2=\left(\begin{array}{cc}
0 & \sin\phi\\
e^{i\xi}\sin\theta & 0
\end{array}\right).
\end{eqnarray}
Here $\theta,\phi,\xi$, and $\eta$ are all real numbers. Apparently, $\Lambda^2$ is an incoherent channel. However, $\Lambda^1$ is \emph{not} an incoherent channel unless $\sin\phi\cos\phi\sin\theta\cos\theta=0$. If this condition is not satisfied, both $E_1^1$ and $E_2^1$ have three nonzero elements, and any linear combination of the two Kraus operators is not incoherent. Recalling that any other Kraus decomposition $\{F_i^1\}_{i=1}^d$ of $\Lambda^1$ is related to $\{E_1^1,E_2^1\}$ by a $d$-dimension unitary transformation $[u_{ij}]_{i,j=1}^d$ as $F_i^1=u_{i1}E_1^1+u_{i2}E_2^1$, (and hence $F_i^1$ are not incoherent), we conclude that $\Lambda^1$ is not an incoherent operation when $\sin\phi\cos\phi\sin\theta\cos\theta\neq0$.

Our result shows that even for the simplest qubit case, there exist non-coherence-generating channels which are not incoherent operations. It implies that the resource theory of coherence based on the incoherent operations is irreversible for general qubit states. This is consistent with the property (C5) in Sec. II. From (C5), $C_f(\rho_2)= C_r(\rho_2)$ holds for a qubit state $\rho_2$ iff $\rho_2$ is pure or incoherent. In other words, if $\rho_2$ is a coherent mixed state, then $C_f(\rho_2)>C_r(\rho_2)$ and hence the coherence transformation is irreversible.

\section{Conclusion}
The non-coherence-generating channels, and their effect on different coherence measures, have been investigated. The relative entropy of coherence for any finite-dimension state, as well as the coherence of formation for any qubit state, are monotonically decreasing under the non-coherence-generating channels. However, the monotonicity of coherence of formation under NCs does not hold for higher-dimension states. Since $C_f$ is always decreasing under the incoherent operations, the nonmonotonicity of $C_f$ under a NC serves as a criterion that the NC is not an incoherent operation.

We define the coherence increasing power of a channel as its ability to increase the coherence measured by either $C_r$ or $C_f$. An interesting effect called the superadditivity of coherence increasing power is proved. Namely, when two channels are applied parallel to a composed system, they may cause larger increase of coherence than used individually. An example is also presented where a qubit NC (which never increases $C_f$ of a single-qubit state) increases $C_f$ of a two-qubit state when applied to one of the two qubits. This property can be employed as a criterion for judging whether a qubit NC is incoherent.

The Kraus presentations of all rank-2 qubit NC channels are derived. Besides ICs, we find a class of NC channels which does not have a Kraus decomposition where all of the Kraus operators are incoherent. This is direct evidence that the IC is a strict subset of the NC.

\begin{acknowledgments}
XH thanks Zi-Wen Liu for stimulating discussions. This work was supported by National Natural Science Foundation of China under Grant No. 11504205, the Fundamental Research Funds of Shandong University under Grant No. 2014TB018, and the National Key Basic Research Program of China under Grant No. 2015CB921003.
\end{acknowledgments}


\section*{Appendices}
\subsection{Proof of Theorem 1.}
By definition, we have
\begin{eqnarray}
C_r(\rho)&=&S(\rho||\Delta(\rho))\nonumber\\
&\geq& S\left(\Lambda^{\mathrm{NC}}(\rho)||\Lambda^{\mathrm{NC}}(\Delta(\rho))\right)\nonumber\\
&\geq& \min_{\sigma\in\mathcal I}S\left(\Lambda^{\mathrm{NC}}(\rho)||\sigma\right)=C_r(\Lambda^{\mathrm{NC}}(\rho)).
\end{eqnarray}
The first inequality is due to the monotonicity of relative entropy under quantum operations, and the second one is because $\Lambda^{\mathrm{NC}}(\Delta(\rho))$ is still an incoherent state but may not be the optimal one in the minimization.

\subsection{Proof of Eq. (\ref{eq:dis}) from Theorem \ref{mono_r}.} Suppose we distill $m$ copy of $\Phi_2$ from $n$ copy of $\rho$ using coherence non-generating operations. In the $n\rightarrow\infty$ limit, the distillation fidelity approaches unity, i.e., $\Phi_2^{\otimes m}=\Lambda^{\mathrm{NC}}(\rho^{\otimes n})$. Then we get
\begin{eqnarray}
m&=&C_r(\Phi_2^{\otimes m})=C_r(\Lambda^{\mathrm{NC}}(\rho^{\otimes n}))\nonumber\\
&\leq&C_r(\rho^{\otimes n})=nC_r(\rho),
\end{eqnarray}
where the first and last equality is from the additivity of $C_r$ under tensor products, and the inequality is from Theorem \ref{mono_r}. So the asymptotical distillable rate $\frac{m}{n}\leq C_r(\rho)$ and hence $C_{dis}^{\mathrm{NC}}(\rho)\leq C_r(\rho)$. On the other hand, $C_{dis}^{\mathrm{NC}}(\rho)\geq C_r(\rho)$, because $\mathcal{IC}\subset\mathcal{NC}$ and $C_{dis}^{\mathrm{IC}}(\rho)=C_r(\rho)$. Therefore, we arrive at Eq. (\ref{eq:dis}).

\subsection{The proof of Lemma 1.}
Let $\{p_k,|\Psi_k\rangle\}$ and $\{p_k,|\psi_k\rangle\}$ be the pure state decompositions of $\rho_{d\times d}$ and $\rho_d$ respectively. Since $\rho_{d\times d}$ is a maximally correlated state, its pure state components must also be a maximally correlated state $|\Psi_k\rangle=\sum_i\lambda^k_i|ii\rangle$. Hence there is a one-to-one correspondence between $\{p_k,|\Psi_k\rangle\}$ and $\{p_k,|\psi_k\rangle\}$ with $|\psi_k\rangle=\sum_i\lambda^k_i|i\rangle$, which satisfies $S(\mathrm{tr}_B(\Psi_k))=S(\Delta(\psi_k))$. By definition, $E_f(\rho_{d\times d})=\min\sum_kp_kS(\mathrm{tr}_B(\Psi_k))$ and $C_f(\rho_d)=\min\sum_kp_kS(\Delta(\psi_k))$, and we arrive at Eq. (\ref{coh_ent}).

\subsection{The proof of Theorem 2.}
From Lemma 2, $C_f(\rho_2)$ is a monotonic function of $C_{l_1}$, and thus Eq. (\ref{mono_f2_e}) is equivalent to
\begin{equation}
C_{l_1}(\Lambda^{\mathrm{NC}}_2(\rho_2))\leq C_{l_1}(\rho_2).\label{mono_l1_e}
\end{equation}
Because $C_{l_1}(\varrho_2)=C_{tr}(\varrho_2)$ and $\arg\min_{\sigma\in\mathcal I}\|\varrho_2-\sigma\|_{tr}=\Delta(\varrho_2)$ for any single-qubit state, we have
\begin{eqnarray}
C_{l_1}(\Lambda^{\mathrm{NC}}_2(\rho_2))&=&\|\Lambda^{\mathrm{NC}}_2(\rho_2)-\Delta(\Lambda^{\mathrm{NC}}_2(\rho_2))\|_{tr}\nonumber\\
&\leq&\|\Lambda^{\mathrm{NC}}_2(\rho_2)-\Lambda^{\mathrm{NC}}_2(\Delta(\rho_2))\|_{tr}\nonumber\\
&\leq&\|\rho_2-\Delta(\rho_2)\|_{tr}=C_{l_1}(\rho_2).
\end{eqnarray}
Here the first inequality is because $\Lambda^{\mathrm{NC}}_2(\Delta(\rho_2))$ is an incoherent state but may not be the nearest one to $\Lambda^{\mathrm{NC}}_2(\rho_2)$, and the second one is from the contractility of trace distance under CPTP map. This completes the proof.

\subsection{Proof of Theorem \ref{mono_f}}
We employ the method of proof by contradiction and assume $C_f(\Lambda^{\mathrm{NC}}(\rho))\leq C_f(\rho),\ \forall\Lambda^{\mathrm{NC}}\in\mathcal{NC},\rho$. Consider a coherence formation protocol, which prepare $n$ copies of $\rho$ out of $m$ copies of $\Phi_2$. In the $n\rightarrow\infty$ limit, the fidelity can reach unity, i.e., $\rho^{\otimes n}=\Lambda_{\mathrm{NC}}(\Phi_2^{\otimes m})$. By the above assumption, we have
\begin{eqnarray}
m&=&C_f(\Phi_2^{\otimes m})\nonumber\\
&\geq& C_f(\Lambda^{\mathrm{NC}}(\Phi_2^{\otimes m}))=C_f(\rho^{\otimes n})\nonumber\\
&=&nC_f(\rho).
\end{eqnarray}
The equations in the first and third lines come from the additivity. Then for any coherence formation protocol, the ratio $\lim_{n\rightarrow\infty}\frac{m}{n}\geq C_f(\rho)$, which is strictly greater than $C_r$ for states which are not in the form of Eq. (\ref{re_state}). It means that, by the assumption that $C_f$ is monotonically decreasing under non-coherence-generating channels, the irreversible state $\rho$ exists such that $C_{cos}^{\mathrm{NC}}(\rho)>C_{dis}^{\mathrm{NC}}(\rho)$. This is wrong because we have considered the maximal set of coherence non-generating operations and hence the coherence distillation and formation should be reversible for all states. This completes the proof.

\subsection{Detailed calculation in the example of superadditivity}
We show that the coherence of formation for the output state $\rho_{out}=\iden_2\otimes\Lambda^2_2(\Phi^+)$ is strictly larger than 1. 
Since $\rho_{out}$ is supported on the subspace $\mathcal S_2$ spanned by $|v_1\rangle$ and $|v_2\rangle$, any pure state component of $\rho_{out}$ should also be supported on $\mathcal S_2$, and hence can be written as $|\psi_i\rangle=\cos\theta|v_1\rangle+\sin\theta e^{i\phi}|v_2\rangle$. Let $a_{\pm}=|\cos\theta\sin\frac{\pi}{8}\pm\sin\theta\cos\frac{\pi}{8}e^{i\phi}|^2$ and $b_{\pm}=|\cos\theta\cos\frac{\pi}{8}\pm\sin\theta\sin\frac{\pi}{8}e^{i\phi}|^2$, and we have
\begin{eqnarray}
S(\Delta(\psi_i))&=&-\frac{a_+}{2}\log_2\frac{a_+}{2}-\frac{a_-}{2}\log_2\frac{a_-}{2}\nonumber\\
&&-\frac{b_+}{2}\log_2\frac{b_+}{2}-\frac{b_-}{2}\log_2\frac{b_-}{2}\nonumber\\
&=&1+\frac12h(a_+)+\frac12h(a_-)>1.\label{eq:ex}
\end{eqnarray}
The last inequality is because the non-negative functions $h(a_+)$ and $h(a_-)$ does not vanish simultaneously. Notice that Eq. (\ref{eq:ex}) holds for any pure state $|\psi_i\rangle$ supported on $\mathcal S_2$, so we have
\begin{eqnarray}
C_f(\rho_{out})&=&\min_{\{p_i,|\psi_i\rangle\}}\sum_ip_iS(\Delta(\psi_i))\nonumber\\
&\geq&\min_{|\psi_i\rangle\in\mathcal S_2} S(\Delta(\psi_i))>1.
\end{eqnarray}
\subsection{Proof of Theorem 4.}
By definition, we have
\begin{eqnarray}
&&P_{C}(\Lambda_{d_1})+P_{C}(\mathcal E_{d_2})\nonumber\\
&=&\sup_{\rho_1\in\mathcal H_{d_1}}[C(\Lambda_{d_1}(\rho_1))-C(\rho_1)]\nonumber\\
&&+\sup_{\rho_2\in\mathcal H_{d_2}}[C(\mathcal E_{d_2}(\rho_2))-C(\rho_2)]\nonumber\\
&=&\sup_{\rho_1,\rho_2}C(\Lambda_{d_1}\otimes\mathcal E_{d_2}(\rho_1\otimes\rho_2))-C(\rho_1\otimes\rho_2)\nonumber\\
&\leq&P_C(\Lambda_{d_1}\otimes\mathcal E_{d_2}).
\end{eqnarray}
The second equality is from the additivity of $C_r$ and $C_f$.

\end{document}